\documentclass[12pt]{amsart}
\usepackage[cp1251]{inputenc}
\usepackage[english,russian]{babel}
\usepackage{amsaddr}
\usepackage{amsmath}
\usepackage{amssymb}
\usepackage{amsfonts}
\usepackage{epsfig}
\usepackage{srcltx}
\usepackage{cite}
\usepackage{float}
\usepackage[ a4paper,  mag=1000, includefoot,  left=2cm, right=2cm, top=2cm, bottom=2cm, headsep=1cm, footskip=1cm ]{geometry}

\begin{document}

\thispagestyle{empty}

\title[]{Influence of small dispersion on self-focusing in spatially one-dimensional case}

\author{Suleimanov B. I.}

\address{Institute of Mathematics, Ufa Scientific Center, 112 Chernyshevsky str., Ufa 450008, Russia}

\email{bisul@mail.ru}

\maketitle {\small
\begin{quote}


\end{quote}
\begin{quote}
\noindent{\bf Abstract.}
The effect of the small dispersion on the self-focusing of solutions of the equations of nonlinear geometric optics in one-dimensional case is investigated. In the main order this influence is described by means of the universal special solution of the nonlinear Schrцdinger equation, which is isomonodromic. Analytic and asymptotic properties of this solution are described.
\medskip

\end{quote} }

\newpage

\begin{center}
{\bf Влияние малой дисперсии на самофокусировку в пространственно одномерном случае}

{Сулейманов Б. И. }

{\it Институт математики с ВЦ УНЦ РАН, 450008 Уфа, Россия, ул. Чернышевского, 112}

{\it bisul@mail.ru}

\end{center}
  
\noindent{\bf Аннотация.}
Исследуется влияние малой дисперсии на самофокусировку решений уравнений нелинейной геометрической оптики в пространственно одномерном случае. В главном порядке это влияние описывается с помощью универсального специального решения 
нелинейного уравнения Шредингера, являющегося изомонодромным. Описаны аналитические и асимптотические свойства этого универсального решения.

\medskip


\bigskip

{\bf1.} 
Решениям пространственно одномерных уравнений нелинейной геометрической оптики (НГО) 
\begin{align}\rho'_{T}+(\rho v)'_X=0,\quad v'_{T}+vv'_X-4\rho'_X=0,\label{NGO}\end{align} часто используемых для описания самых разных явлений в неустойчивых средах, присуще~\cite{bib:Gsh}--\cite{bib:Zhd} их разбиение на отдельные самостягивающиеся сгустки, разделенные 
промежутками нулевой интенсивности $\rho=0$. В \S26, \S28 и \S32 монографии Жданова и Трубникова~\cite{bib:Zhd} содержится важный вывод о том, что в ситуации <<общего положения>> такие сгустки {\it самофокусируются} и что, по -- видимому, cингулярности cоответствующих решений системы НГО в окрестностях фокальных точек 
$T=T_f$, $X=X_f$ в главном порядке задаются автомодельным решением сиcтемы ~(\ref{NGO}) c 
($s=(X-X_f)(T-T_f)^{-2/3}$, $a$ -- постоянная)
\begin{align}
\rho(T,X)&=(T-T_f)^{-2/3}[\frac{a}{2}-\frac{s^2}{36}]\label{avtomp}\end{align}
($\rho(T,X)$=0 вне стягивающегося при $T\to T_f$ промежутка 
$|X-X_f| \leq \sqrt{18a}(T-T_f)^{2/3}$). 

Данная статья посвящена описанию влияния малой дисперсии на такие процессы самофокусировки, необходимость изучения которого, в частности, отмечена в заключительном абзаце \S32 монографии ~\cite{bib:Zhd}. 

{\bf2.} Эталонный пример уравнения, на решениях которого будет изучаться это влияние, представляет нелинейное уравнение Шредингера (НУШ)
\begin{align}-i\varepsilon G'_{T}=\varepsilon^2G''_{XX}+2|G|^2 G\label{Nsh} \qquad(\varepsilon<<1).
\end{align}

Приближение НГО к решениям данного уравнения возникает следующим образом: 
подстановка \begin{align}G = \rho^{1/2}\exp\left(\frac{i\Phi}{\varepsilon}\right)\label{quasi}\end{align} в НУШ ~(\ref{Nsh}) дает систему 
\begin{align}\rho '_{T}+ 2(\rho \Phi'_X)'_ X = 0,\quad\Phi'_{T} + (\Phi'_X)^2 - 2\rho = \varepsilon^2\frac{( \sqrt{\rho })''_{XX}}{\sqrt \rho } Х,\label{smngo}\end{align}
бездисперсионный предел которой ($\varepsilon=0$) 
\begin{align}\rho '_{T}+ 2(\rho \Phi'_X)'_ X = 0,\quad    \Phi'_{T} + (\Phi'_X)^2 - 2\rho =0, \label{pngo}\end{align}
после дифференцирования по переменной $X$ второго ее уравнения 
сводится к системе НГО~(\ref{NGO}) c $v= 2\Phi'_X$.

Из автомодельности решения этой системы с компонентой~(\ref{avtomp}) следует, что вторая его компонента имеет вид $v(T,X)=(T-T_f)^{-1/3}2s/3$. А значит, соответствующие самофокусирующиеся сгустки, описываемые решениями системы~(\ref{pngo}), при $T\to T_f-0$ и $X\to X_f$ на некотором стягивающемся к точке $X=X_f$ промежутке ненулевой интенсивности имеют асимптотики ($\Phi_*$ -- поcтоянная) \begin{equation}\begin{aligned}
\Phi(T,X)=&\Phi_*+(T-T_f)^{1/3}[3a+\frac{s^2}{6}]+\dots,\\
\rho(T,X)=&(T-T_f)^{-2/3}[\frac{a}{2}-\frac{s^2}{36}]\dots.\label{polst}
\end{aligned}\end{equation}

{\bf 3.} Действуя cогласно методу согласования~\cite{bib:Il'}, с целью учета поправочного действия правой части системы~(\ref{smngo}) на самофокусировку решений~(\ref{pngo}) осуществим следующие действия: 

1) сделаем такие замены 
\begin{align}\label{scal}X-X_f=x \varepsilon^{\alpha}, T-T_f=t \varepsilon^{3\alpha/2}, G=\exp\left(\frac{i\Phi_*}{\varepsilon}\right)\frac{q}{\varepsilon^{ \gamma}}, \end{align}
 чтобы в образе~(\ref{Nsh}) -- уравнении 
$$ -i\varepsilon^{1-3\alpha/2}q_t=\varepsilon^{2-2\alpha}q_{xx}+2\varepsilon^{-2\gamma}|q|^2q$$
 -- все члены были одного порядка. Таким образом, получаем, что в заменах (\ref{scal})  $\alpha=2$, $\gamma=1$, 
и что решение уравнения~(\ref{Nsh}) в окрестности точки фокуса $(T_f,X_f)$ в главном порядке задается формулой 
\begin{align}\label{gl} G=\frac{q(t,x)}{\varepsilon}+...,\end{align}
где $q(t,x)$ есть 
гладкое при всех $t$ и $x$ решение $q(t,x)=r(t,x)\exp{(i\varphi(t,x))}$ фокусирущего НУШ
\begin{align}\label{nush} -iq_t=q_{xx}+2|q|^2q, \end{align}
уже {\it независящее} от малого параметра $\varepsilon$; 

2) из условия согласования с асимптотиками~(\ref{polst}) выведем, что это глобально гладкое решение НУШ~(\ref{nush}) при $t\to -\infty$ вне двух ветвей кривой 
\begin{align}\label{parab}x^2=18at^{4/3}\end{align}
стремится к нулю, а внутри их его амплитуда $r(t,x)$ и фаза $\varphi(t,x)$ в главном задаются формулами 
\begin{equation}\begin{aligned}
\varphi(t,x)=&t^{1/3}[3a+\frac{s^2}{6}]+\dots,\\
r(t,x)=&t^{-2/3}[\frac{a}{2}-\frac{s^2}{36}]^{1/2}+\dots.\label{avtom}
\end{aligned}\end{equation}

В конце раздела {\bf7} будет показано, что амплитуда $r(t,x)$ {\it четна} по переменной $t$,
и что разность $\varphi(t,x)-\varphi_0$ межу фазой $\varphi(t,x)$ и некоторой постоянной $\varphi_0$ по $t$  {\it нечетна}.

А значит,
 вне малой окрестности точки фокуса $(T_f,X_f)$ после критического момента $T=T_f$ при некоторых конечных значениях $|X-X_f|^2+|T-T_F|^2$ соответствующие решения уравнения с малой дисперсией~(\ref{Nsh}) вновь описываются квазиклассическим приближением ~(\ref{quasi}), определяемым решениями системы~(\ref{pngo}). В данной области эти решения~(\ref{pngo}) некоторое время будут описывать расфокусирующиеся сгустки, асимптотика которых на некотором сужающемся при $T\to T_f+0$ промежутке ненулевой амплитуды также описывается формулами~(\ref{polst}). Далее естественно ожидать, что по истечении некоторого времени эти сгустки вновь начнут сужаться и самофокусироваться. (Правда, в принципе, у соответствующих решений системы НГО за точками фокуса возможны ~\cite{bib:Gsh}--~\cite{bib:Zhd} и другие особенности -- см. также~\cite{bib:Dub},~\cite{bib:Tovb} и ссылки в последней из этих работ.)

{\bf 4.} Специальное решение $q(t,x)$ НУШ~(\ref{nush}), введенное в рассмотрение в предыдущем разделе, имеет довольно универсальный характер. 

Рассмотрим, например, cтационарное самовоздействие модулированной волны, описываемое
~\cite{bib:Vinsr} нелинейным уравнением 
Гельмгольца
\begin{align}\label{Helm}E''_{YY}+E''_{ZZ}+\frac{\omega^2}{c^2}\epsilon(|E|^2)E=0\quad(\frac{\omega}{c}>>1),\end{align} в котором при $|E|\to 0$ ($\epsilon_n$ -- постоянные)
\begin{align}\label{Diel}
\epsilon(|E|^2)=\epsilon_0+\epsilon_2|E|^2+\epsilon_4|E|^4+\sum\limits_{j=3}^{\infty}\epsilon_{2j}|E|^{2j}.
\end{align}
Представим $E$ в виде ($\mu<<1$, $k=\omega \epsilon_0/{c}>>1$)
\begin{align}\label{sme}E=\sqrt{\mu} W(X,T,\Theta,\mu,k), \end{align}
где $$X=\sqrt{\mu}Y,\quad T=\frac{\mu Z}{2},\quad\Theta=k Z.$$ Подстановка~(\ref{sme}) в~(\ref{Helm}), (\ref{Diel}) дает уравнение
\begin{equation}\begin{aligned}k^2W''_{\Theta\Theta}+\mu kW''_{\Theta T}+\mu^2\frac {W''_{TT}}{4}+\mu W''_{XX}+k^2W\left(1+\right.\\\left.
+\mu\frac{\epsilon_2}{\epsilon_0}W^2+\mu^2\frac{\epsilon_4}{\epsilon_0}W^4+\sum\limits_{j=3}^{\infty}\mu^j\frac{\epsilon_{2j}}{\epsilon_0}W^{2j}\right)=0,\label{smw}\end{aligned}\end{equation}
решение которого будем искать в виде ряда
\begin{align}\label{stepm}W=W_0(X,T,\Theta,k)+\mu W_1(X,T,\Theta,k)+\dots.\end{align}

Этот анзац для решения уравнения~(\ref{smw}) дает рекуррентную последовательность обыкновенных дифференциальных уравнений (ОДУ) по независимой переменной $\Theta$ на коэффиценты $W_n$ ряда~(\ref{stepm}): 
$$k^2[(W_0)''_{\Theta\Theta}+W_0]=0,$$
$$k^2[(W_1)''_{\Theta\Theta}+W_1]=-k(W_0)''_{\theta T}-(W_0)''_{XX}-k^2\frac{\epsilon_2}{\epsilon_0}(W_0)^3,$$
$$....$$
Из вида общего решения первого из этих ОДУ
\begin{align}\label{Wood}
W_0=H(T,X,K)\exp{(i\Theta)}+H^{*}(T,X,K)
\exp{(-i\Theta)}
\end{align}
и условия отсутствия секулярных членов (пропорциональных $\Theta\exp{(\pm\Theta)}$) в решении $W_1$ второго ОДУ данной последовательности следует, что комплексная амплитуда $H(T,X,K)$ главного члена~(\ref{Wood}) ряда~(\ref{stepm}) после растяжения
$G(T,X,\varepsilon)=\sqrt{3\epsilon_2/(2\epsilon_0)}H$
будет определяться решением НУШ
~(\ref{Nsh}) 
c малым параметром $\varepsilon=k^{-1}$. 
Предположим, что к этим решениям $G$ применимо cамофокусирующееся квазиклассическое приближение НГО, описанное в разделах {\bf 1}, {\bf2}. Тогда при условии, что параметры $\mu$ и $k$ cвязаны соотношением 
$\mu k^2<<1,$
в главном по $k$ порядке данные решения $G$ в окрестностях точек фокуса $(T_f,X_f)$ 
тоже
задаются формулой~(\ref{gl}). 

В самом деле, растяжение
$W=kB,$
согласующееся с формулой ~(\ref{gl}), и замены независимых переменных~(\ref{scal}) переводят~(\ref{smw}) в уравнение 
\newpage
$$B''_{\Theta\Theta}-B+\mu k^2\left(B''_{\Theta t}+ B''_{xx}+\frac{\epsilon_2}{\epsilon_0}B^3\right)=$$
$$=-\mu^2k^4\left(\frac{B''_{tt}}{4}+
\frac{\epsilon_4}{\epsilon_0}B^5
+\sum\limits_{j=1}^{\infty}(\mu k^{2})^j
\frac{\epsilon_{2(j+2)}}{\epsilon_0} B^{2j+5}\right).
$$
Применяя далее стандартную процедуру разложения решения последнего уравнения в ряд\linebreak $B=\sum_{j=0}^{\infty}(\mu k^{2})^j B_j(t,x,\Theta)$, после уничтожения секулярных членов при определении $B_1$ приходим к выводу о справедливости утверждения, сформулированного в конце предыдущего абзаца.

{\bf 5.} В окрестности точки фокуса $(T_f,X_f)$ соотношениями~(\ref{polst}) описывается и известный самофокусирующийся импульс Таланова~\cite{bib:Tal}, 
фактически соответствующий~\cite{bib:Gsh} точным решениям системы~(\ref{pngo}) 
\begin{equation}\begin{aligned}
\Phi(T,X)&=\delta(T)+g(T)X^2,\\
\rho(T,X)&=\frac{\delta_T}{2}+X^2\left[\frac{g'_T}{2}+2g^2(T)\right],
\end{aligned}\label{Tal}\end{equation}
где, как легко видеть, $$g''_{TT}+20gg'_T+48g^3=0,\qquad \delta''_{TT}+4g\delta'_T=0.$$
(Вне интервала $|X|<\left[-\delta_T/(g'_T+4g^2)\right]^{1/2}$ этот импульс тождественно равен нулю.)

Так что в случае применимости приближения НГО к решениям уравнения~(\ref{Nsh}), описываемого такими самофокусирующимися решениями ~(\ref{Tal}), поведение приближаемых решений $G(T,X,\varepsilon)$
 будет, понятно, также задаваться формулой~(\ref{gl}). 

З а м е ч а н и е 1. Вывод cтатьи~\cite{bib:KMsh} о неприменимости этого приближения НГО для решений НУШ к ситуации уравнения~(\ref{Nsh}) с параметром $\varepsilon<<1$, напрямую не относится. (Попытка свести эту ситуацию к той же, что в ~\cite{bib:KMsh} приводит к тому, что вместо 1 в неравенстве $N>>1$ из раздела {\bf 2}~\cite{bib:KMsh} будет стоять малый параметр.) Однако вопрос о применимости к решениям~(\ref{Nsh}) данного приближения Таланова все же пока надо признать остающимся открытым.

{\bf 6.} Описанное выше универсальное решение $q(t,x)$ НУШ~(\ref{nush}) 
является представителем множества специальных решений интегрируемых методом обратной задачи рассеяния (МОЗР) уравнений -- аналогов так называемых специальных функций волновых катастроф~\cite[Гл.6, \S4]{bib:Fed}. Общая теория таких решений, изучение которой 
было начато Китаевым в публикации в Записках ЛОМИ~\cite{bib:Kit} (независимо частный случай одного подобного решения рассматривался автором данной статьи в том же номере Записок ЛОМИ ~\cite{bib:Loms}), позднее получила дальнейшее развитие и уже довольно многочисленные применения~\cite{bib:Matz}--~\cite{bib:GGS}.

З а м е ч а н и е 2. К этим же решениям обычно приводит и использование методики Дубровина~\cite{bib:DubB}, исходящей из приближенных симметрий нелинейных уравнений с малым параметром~\cite{bib:BGI}.

Как и другие представители данного множества, это решение $q$ НУШ~(\ref{nush}) относится к классу {\it изомонодромных} (ИДМ)~\cite{bib:Its}: наряду с сиcтемами линейных уравнений МОЗР~\cite{bib:Zsh}
($\lambda$ -- параметр) 
\begin{align}\label{mozx}\Psi'_x=i\begin{pmatrix}-\lambda&q\\q^{*}&\lambda\end{pmatrix}\Psi,\end{align}
\begin{align}\label{mozt}\Psi'_t=\begin{pmatrix}-i(2\lambda^2-|q|^2)&2i\lambda q-q_x\\2i\lambda q^{*}+q^{*}_x&i(2\lambda^2-|q|^2)\end{pmatrix}\Psi,\end{align}
их соответствующее совместное решение $\Psi$ удовлетворяет также системе линейных ОДУ 
\begin{align}\label{mozl}\Psi'_{\lambda}=A(\lambda,t,x)\Psi,\end{align}
где матрица $A(\lambda ,t,x)$ рациональна по переменной $\lambda$.

Исходя из вида переменной $s=xt^{-2/3}$ в асимптотике~(\ref{avtom}) этого решения $q(t,x)$ НУШ~(\ref{nush}) и выводов из~\cite{bib:Kit} -- \cite{bib:GGS}, естественно сопоставить $q(t,x)$ интеграл 
\begin{align}\label{furfe} I(t,x)=\int_L\lambda^m\exp{i(\lambda x-\lambda^2 t+\delta \lambda^{-1})}d\lambda. \end{align}
Здесь $\delta$, $m$ -- постоянные, а $L$ -- такой контур, что интегрирование по частям этого интеграла  не дает вкладов от внеинтегральных членов. (Интеграл Фурье~(\ref{furfe}) удовлетворяет линейной части НУШ~(\ref{nush}) $-iI_t=I_{xx}$ и его асимптотика при больших значениях $|t|$ описывается~\cite{bib:Fed} как раз после замены $x=s t^{2/3}$.)

А уже из вида подинтегральной экспоненты этого линейного аналога $q(t,x)$ с учетом гладкости $q(t,x)$ и вида 
 систем~(\ref{mozx}), (\ref{mozt}) согласно общей идеологии~\cite{bib:Kit} (и более конкретно-применительным рассуждениям и выкладкам из~\cite{bib:Matz}) вытекает, что в ОДУ~(\ref{mozl}) 
\begin{align}A(\lambda,t,x)=\sum\limits_{j=-2}^{1}A_j(t,x)\lambda^j,\label{isoA}\end{align}
где
$$A_1=4i t \begin{pmatrix}-1&0\\0&1\end{pmatrix}, \qquad A_0=\begin{pmatrix}-ix&4itq\\4itq^*&ix\end{pmatrix}, $$
$$A_{-1}= \begin{pmatrix}2it|q|^2&ixq-2tq_x\\ixq^*+2tq^*_x&-2it|q|^2\end{pmatrix}, $$
$$A_{-2}=
[t(q_xq^*-q^*_xq)-$$
$$-\frac{i}{2}(x|q|^2
+\int\limits_0^x|q|^2(t,\zeta)d\zeta+\nu)]\begin{pmatrix}-1&0\\0&1\end{pmatrix}+$$
$$+\begin{pmatrix}0&-tq'_t-(xq)'_x/2\\t(q^*)'_t+
(xq^*)'_x/2&0\end{pmatrix}. $$

Следствием совместности на данном изомонодромном решении $q(t,x)$ cистем линейных ОДУ~
(\ref{mozx}) и~(\ref{mozl}),~(\ref{isoA}) является тот факт, что наряду с НУШ~(\ref{nush}) $q(t,x)$ удовлетворяет и нелинейному ОДУ 
\begin{equation}\begin{aligned}it[q_{xxx}+6|q|^2q_x]+\frac{xq_{xx}}{2}+q_x+x|q|^2q+\\
+2(\nu+ \int\limits_0^x|q|^2(t,\zeta)d\zeta)q=0.\label{odex}\end{aligned}\end{equation}
Аналогично, из совместности систем линейных ОДУ~(\ref{mozt}) и~(\ref{mozl}) cледует, что $q(t,x)$ удовлетворяет также нелинейному ОДУ по независимой переменной $t$. При этом  $\det A_{-2}=const$. Исходя из асимптотик~(\ref{avtom}), находим, что в нашем  случае 
\begin{align}\label{integ}
\det A_{-2}=\frac{a^3}{2}.\end{align}

{\bf 7.} Аналогии с поведением совместных решений линейных частей ОДУ~(\ref{odex}) и НУШ ~(\ref{nush}) позволяют прийти к выводу о том, что при $-t>>1$ в области провала интенсивности $|q|^2$, которая расположена вне полукубической параболы~(\ref{parab}), асимптотика $q(t,x)$ должна иметь вид
\begin{align}\label{linas}\frac{1}{t^{1/2}}\sum\limits_{j=1}^3\beta_j \frac{|f_j(s)|^{3/2}\exp{(iH_j(t,s))}}{|f_j(s)^3-d|^{1/2}}+\dots.\end {align}
Здесь
$$H_j=t^{1/3}\left(sf_j(s)-f_j(s)^2+\frac{d}{f_j(s)}\right)+\gamma_j \ln t+\alpha_j(s),$$
$\beta_j$, $\gamma_j$ -- постоянные, а функции $f_j(s)$, как и в асимптотиках некоторых линейных аналогов $q(t,x)$ вида~(\ref{furfe}), вычисляемых согласно методу стационарной фазы~\cite{bib:Fed}, определяются из кубического уравнения 
$s-2f_j(s)-{d}/f^2_j(s)=0.$

Потребовав теперь, чтобы асимптотики~(\ref{linas}) теряли пригодность на полукубической параболе~(\ref{parab}), с необходимостью получаем, что в двух последних выделенных формулах $d=(2a)^{3/2}$. 
 В свою очередь, из этого факта вытекает, что в~(\ref{isoA}) и
~(\ref{odex}) 
\begin{align}\label{nul}\nu=0.\end{align}

При $t=0$ из ОДУ~(\ref{odex}) и равенства~(\ref{nul}) следует справедливость равенства $q'_x(0,0)=0$ и четность по $x$ функции $q(0,x)$. Рассматривая теперь последнюю в качестве начального данного при $t=0$ задачи Коши для НУШ~(\ref{nush}), приходим к выводу о четности нашего решения $q(t,x)$ по переменной $x$ и при всех $t$.

По этой причине при $x=0$ условием совместности уравнений ИДМ~(\ref{mozt}) и~(\ref{mozl}),~(\ref{isoA}) является удовлетворение функцией $q(t,0)$ нелинейному ОДУ
\begin{align} tq''_{tt}+\frac{3q'_t}{2}=2it|q|^2q'_t+i|q|^{2}q.\label{qunt}\end{align}
Все гладкие  решения $q(t,0)=r(t,0)\exp{(i\varphi(t,0))}$ этого ОДУ таковы, что их амплитуда $r(t,0)$ и производная фазы $\varphi'_t(t,0)$ чeтны по $t$. 
А значит, и при всех $x$ амплитуда $r(t,x)$ и производная фазы $\varphi'_t(t,x)$ данного cовместного решения $q(t,x)=r(t,x)\exp{(i\varphi(t,x))}$ НУШ~(\ref{nush}) и ОДУ~(\ref{odex}) 
четны по переменной $t$. Из справедливости НУШ~(\ref{nush}) при $t=0$ выводим, что
$\varphi(0,x)'_xr^2(0,x)=const$. Из этого же соотношения, из ~(\ref{nul}) и из ОДУ~(\ref{odex}) cледует, что $\varphi(x,0)=\varphi_0=const$.

{\bf 8.} Так как $q(0,x)=r(0,x) \exp{(i\varphi_0)},$ то в силу четности 
$q(0,x)$, вида ОДУ~(\ref{odex}) и равенств~(\ref{integ}),~(\ref{nul}) получаем, что $r(0,x)$ при $x\to 0$ раскладывается в ряд
\begin{align}\label{nnul}r(0,x)=(2a^3)^{1/2}-(2a^3)^{3/2}x^2+\dots.\end{align}
Это решение $r(0,x)$ ОДУ~(\ref{odex}) посредством замен 
$$y=2^{3/2}(2a^3)^{1/4}x^{1/2},\quad r(0,x)=2(2a^3)^{1/2}\frac{w'_y}{y}$$
выражается через четное по переменной $y$ решение $w(y)$ частного случая третьего уравнения Пенлеве $w''_{yy}+w'_y/y+\sin{w}=0,$ удовлетворяющее условию $w(0)=-\pi/2$. Из известной~\cite{bib:Nov} асимптотики $w(y)$ при $y\to\infty$  находим, что при $x>>1$ 
$$r(0,x)\approx \frac{h}{x^{3/4}}\sin{\left(2(2a)^{3/4}x^{1/2}+b \ln x+c\right)},$$
где 
$h=(2a)^{3/8}\sqrt{\ln2/{\pi}}$, $b=\ln2/(4\pi)$, $$c=\arg{\Gamma\left(\frac{i\ln2}{2\pi}\right)}-\frac{\pi}{4}\left(1+(\ln2)^2\right)+\frac{\ln2 \ln(2a^3)}{8\pi}.$$

{\bf 9.} При $x=0$ из четности по $x$ нашего измонодромного решения $q(x,t)$ cледует, что пара  совместных на решениях ОДУ~(\ref{qunt}) линейных уравнений ИДМ~(\ref{mozt}) и~(\ref{mozl}),~(\ref{isoA}), cводится к паре уравнений ИДМ из классической работы~\cite{bib:Gar}, условием совместности которой является удовлетворение функцией 
$$ \xi(t)=\frac{8t}{a^3}\left[t \varphi'_t(t,0)r^2(t,0)+\frac{i}{2}(t r^2(t,0))'_t \right]$$
следующему случаю третьего уравнения Пенлеве
\begin{align}\label{dept}\xi''_{tt}=\frac{(\xi'_t)^2}{\xi}-\frac{\xi'_t}{t}-\frac{a^3\xi^2}{4t^2}+
\frac{2i}{t}-\frac{16}{\xi}.\end{align}
Из соотношения ~(\ref{nnul}) и чeтности функций $r(t,0)$, $\varphi_t(t,0)$, можно, воспользовавшись результатами статьи~\cite{bib:Kitv}, в явном виде выписать асимптотику
$\xi(t)$ для больших значеий $-t$. А из последней уже вытекает заключение о том, что внутри полукубической параболы~(\ref{parab}) на фоне приближения НГО, задаваемых правыми частями~(\ref{avtom}), происходят двухфазовые малоамплитудные колебания 
с двумя частотами вида
$$t^{1/3}3h_j(s)\sqrt{h_j(s)^2-2a+s^2/9}+\mu_{j} \ln t,$$
где $\mu_{j}$ -- постоянные, а
$$h^2_j(s)=3a-\frac{s^2}{24}\pm\frac{s}{2}\sqrt{a+s^2/144}.$$
{\bf 10}. Двухфазовость этих колебаний и трехфазовость колебаний в асимптотике~(\ref{linas}) качественно соответствуют выводам работ~\cite{bib:Jks}, ~\cite{bib:KTS}, в которых описано влияние малой дисперсии на процессы провального самообострения импульсов приближения НГО. 

Асимптотика $q(t,x)$ в окрестности правой из кривых~(\ref{parab}) описывается (также подобно тому как в ~\cite{bib:Jks}, ~\cite{bib:KTS}) в терминах комплексных решений ОДУ $\omega''_{zz}+(|\omega|^2+z)\omega=0$ по переменной $z=(s-3\sqrt{2a})t^{2/9}$, которые стремятся к нулю при $z \to \infty$, и для которых при $z \to -\infty$ $|\omega(z)|\approx \sqrt{-z/2}$. Формулой $f=(\ln\omega)'_z$ эти его решения задают~\cite{bib:Jim} комплексные сингулярные решения второго уравнения Пенлеве $f''_{zz}=2f^3+2zf+\kappa$, где $\kappa=\omega_z\omega^*-\omega_z^*\omega-1$.

З а м е ч а н и е 3. 
Аналогичным образом, вероятно (см. абзац после формулы (4.18) в статье~\cite{bib:ZsK}), подобная переходная область может быть описана и в пространственно многомерных случаях.

{\bf 11.} Автор выражает глубокую благодарность А.\,М. Камчатнову, по настоятельному совету которого, собственно, и было проведено исследование, являющееся предметом данной статьи. Он также признателен Д.\,П. Новикову за указание на то, что при $x=0$ и $q_x(t,0)=0$ 
уравнения (\ref{mozt}),~(\ref{mozl}),~(\ref{isoA}) cводятся к паре уравнений ИДМ для третьего уравнения Пенлеве
~(\ref{dept}) из ~\cite{bib:Gar}, и А.\,В. Китаеву за консультации по поводу свойств соответствующего решения $\xi(t)$ этого уравнения (в особенности, за замечание о симметрии $\xi(t)$ по переменной $t$). Весьма полезными при написании данной статьи были обсуждения части ее результатов с Е.\,А. Кузнецовым и А.\,В. Михайловым.

\end{document}